\documentclass[12pt]{article}
\usepackage[utf8]{inputenc}
\usepackage{amsmath, amssymb}
\usepackage{graphicx}
\usepackage{float}
\usepackage{xcolor} 
\usepackage{authblk}

\title{Structural and thermodynamic stability of hexagonal-diamond $\text{Si}_{1 - x - y} \text{Ge}_{x} \text{B}_{y}$ alloys}

\author[1]{Marc Túnica}
\author[2]{Francesca Chiodi}
\author[1]{Michele Amato\thanks{michele.amato@universite-paris-saclay.fr}}

\affil[1]{Université Paris-Saclay, CNRS, Laboratoire de Physique des Solides, 91405 Orsay, France}
\affil[2]{Université Paris-Saclay, CNRS, Centre de Nanosciences et de Nanotechnologies, 91120, Palaiseau, France}

\date{}

\begin{document}

\maketitle

\begin{abstract}    
Pushing dopant concentrations beyond the solubility limit in semiconductors—a process known as hyperdoping—has been demonstrated as an effective strategy for inducing superconductivity in cubic-diamond Si and SiGe materials. Additionally, previous studies have reported that several polytypes of Si may exhibit a type-I superconducting state under high pressure. In this work, we employ ground-state Density Functional Theory simulations to investigate the effects of both low and high B doping concentrations on the structural and thermodynamic properties of hexagonal-diamond SiGe alloys, with a systematic comparison to their cubic-diamond counterparts. Our results highlight three key findings: (i) structural analysis confirms that the lattice parameters of SiGeB alloys adhere to a ternary Vegard’s law, consistent with observations in cubic-diamond SiGe alloys. However, at high doping concentrations, B incorporation can locally disrupt the hexagonal symmetry, particularly in the presence of B clustering; (ii) dopant formation energy calculations reveal that B is thermodynamically more stable in the hexagonal phase than in the cubic phase across all Ge concentrations, regardless of the doping level; (iii) mixing enthalpy calculations demonstrate that hyperdoped hexagonal-diamond SiGe alloys are thermodynamically stable across the full range of Ge compositions and that their tendency for hyperdoping is more favorable than that of cubic-diamond SiGe alloys. Taken together, these findings indicate that hyperdoping is experimentally viable in hexagonal-diamond SiGe alloys and, in light of previous evidence, position these materials as a promising platform for the exploration of superconductivity in group IV semiconductors.  
\end{abstract}

%
%
%
%
%

\section{Introduction}
Achieving ultra-doping in covalent semiconductors or insulators can turn them metallic and, pushed further to overcome the equilibrium solubility limit, can lead to superconductivity~\cite{Blase2009, KrienerPRB2008}. 
This phenomenon has been extensively predicted and experimentally observed in group IV materials, including heavily B-, Ga-, or Al-doped cubic-diamond Si (cub-Si)~\cite{Kadas2008, Bustarret2006, Skrotzki2010, Thorgrimsson2020}, cub-Ge~\cite{Kadas2008, Prucnal2019, Herrmannsdorfer2009, Sardashti2021, Strohbeen2023, Porret2023, Steele2024}, cub-SiGe alloys~\cite{Blase2009, Nath2024}, and cub-SiC~\cite{KrienerPRB2008,Ren2007}. In parallel, polytypism is also well known as an alternative route for inducing superconductivity, with the superconductivity critical temperature $T_c$ evolving with the crystal symmetries. For instance, experimental studies have shown superconducting phases in both cubic and hexagonal phases of boron-doped SiC~\cite{KrienerSTAM2008}.
Similarly, the cubic-diamond polytype of silicon (cub-Si) undergoes a superconducting transition under high pressure for multiple phases, the hexagonal-close-packed (hpc) and primitive hexagonal (ph) phases, following the also superconducting $\beta-Sn$ phase, with more than a factor of 2 change in $T_c$ ~\cite{Mignot1986}. This finding gains additional significance in light of recent advances in materials growth, which have demonstrated that the hexagonal-diamond phase of Si (hex-Si), Ge (hex-Ge), and SiGe (hex-SiGe) can be stabilized under ambient pressure and synthesized in the form of nanowires~\cite{TizeiNL2025, Vincent2014,Hauge2017, Hauge2015, GalvaoTizei2020, Tang2017, Fadaly2020,PeetersNC2024}. 
Advanced characterization techniques have revealed that these hexagonal polytypes exhibit distinct electronic and optical properties compared to their cubic-diamond counterparts~\cite{Borlido2021, Rodl2019, GalvaoTizei2020, Keller2023, Xiang2022, Wang2021}. In particular, for Ge concentrations above 65\%, hex-SiGe alloys possess a direct band gap ranging from 0.3 to 0.6 eV, making them light-emitting in the infrared spectrum~\cite{Borlido2021, Fadaly2020, Rodl2019}. This unique combination of a direct band gap—a rare feature among group IV semiconductors—and the evidence for superconductivity in several Si polytypes suggests that hex-SiGe materials hold strong potential as candidates for photonic superconducting devices, or, in the case of $p$-doping, for the investigation of exotic superconductivity resulting from the interaction with the holes' strong spin-orbit coupling.

In this work, we employ total energy Density Functional Theory (DFT) calculations to investigate the structural and thermodynamic stability of B-hyperdoped hex-SiGe alloys. The primary objective is to assess the feasibility of the hyperdoping process in these materials to guide future characterizations in the context of superconductivity. To date, doping in hexagonal-diamond group IV semiconductors has been addressed in only a limited number of studies, primarily focused on pure Si and Ge~\cite{Amato2020, Amato2019, Tunica2024}. Among possible dopants, B is particularly well-suited for hexagonal group IV systems due to its low formation energy, which promotes efficient incorporation into the host lattice~\cite{Amato2020, Amato2019}. With its three valence electrons, B maintains three-fold coordination, rendering it more stable in the hexagonal structure than in the cubic one~\cite{Amato2020, Amato2019}. Moreover, B introduces shallow acceptor states in Si~\cite{Tunica2024} and exhibits high electrical activation in Si and Ge~\cite{Mirabella2008}, further supporting its effectiveness as a dopant. Importantly, B is a particularly promising candidate for doping hex-SiGe alloys beyond the solubility limit, as it interacts weakly with planar defects—an effect attributed to its small atomic radius. This behavior has been demonstrated in the case of I\textsubscript{3} basal stacking faults in hexagonal Si~\cite{Tunica2025}. This observation is particularly relevant, as maintaining a uniform dopant distribution becomes critical in the presence of intrinsic planar defects, which are common in the alloy due to the metastable nature of the hexagonal-diamond phase~\cite{Raffy2002, Fadaly2021, Rovaris2024, Rovaris2025, Vincent2022}. These considerations reinforce the relevance of B as an ideal dopant for advancing the synthesis and functionalization of hex-SiGe alloys.

We begin by analyzing the structural properties of hex-$\text{Si}_{1 - x - y} \text{Ge}_{x} \text{B}_{y}$ alloys and establish that they obey a ternary Vegard’s law, consistent with behavior previously observed in cub-$\text{Si}_{1 - x - y} \text{Ge}_{x} \text{B}_{y}$ alloys~\cite{Nath2024}. Next, we examine single-dopant formation energies and find that B is thermodynamically more stable in the hexagonal phase than in the cubic phase across the entire range of Ge concentrations, irrespective of the doping level. Finally, mixing enthalpy calculations indicate that hyperdoped hex-\(\text{Si}_{1 - x - y} \text{Ge}_{x} \text{B}_{y}\) alloys exhibit slightly positive mixing enthalpies, suggesting they may be thermodynamically stable across the entire composition range. Moreover, their energetic tendency for hyperdoping remains significantly more favorable than that of their cubic-diamond counterparts.

\section{Methodology}
We performed Density Functional Theory (DFT) simulations using the SIESTA code~\cite{Soler2002}. The exchange-correlation energy was treated within the Generalized Gradient Approximation (GGA), using the Perdew–Burke–Ernzerhof (PBE) functional~\cite{Perdew1996}. For Si valence electrons, we employed double-$\zeta$ polarized basis sets, while polarized basis sets were used for Ge and B. Inner core electrons were replaced by norm-conserving pseudopotentials generated using the Troullier–Martins method~\cite{Troullier1991}. Structural optimizations were carried out using the conjugate gradient algorithm, with convergence criteria of 0.01~eV/{\AA} for atomic forces and 0.1~GPa for the stress tensor.

For bulk hex-Si and hex-Ge, structural relaxations were performed using a primitive unit cell containing four atoms and an $8 \times 8 \times 8$ Monkhorst–Pack k-point mesh~\cite{Monkhorst1976}. For the bulk cubic-diamond phase, we adopted the structural data reported in Ref.~\cite{Nath2024}. The converged lattice parameters for Si and Ge in both phases are presented in the first and second columns of Table~\ref{tab:table_1}. These values show good agreement with previous GGA–PBE results~\cite{Baroni1986, Zheng2017}, and, as expected, are slightly overestimated compared to experimental values (shown in parentheses in Table~\ref{tab:table_1}), due to the known tendency of GGA functionals to underestimate binding energies. Bulk crystals were considered a valid approximation for sufficiently large nanowires, as quantum confinement effects are expected to be negligible for diameters exceeding approximately 10~nm. This assumption is supported by previous experimental and theoretical studies~\cite{GalvaoTizei2020, Amato2012, Amato2010, Rurali2010, Kaur2014, David2017,Amato2014}.

SiGe alloys are disordered solid solutions, and an accurate description within DFT would normally require very large supercells to capture the statistical distribution of Si and Ge atoms. To overcome this computational limit, we employed the Special Quasirandom Structures (SQS) method~\cite{Wei1990,Zunger1990}, as implemented in the Alloy Theoretic Automated Toolkit (ATAT)~\cite{VanDeWalle2013}. The SQS approach allows the construction of supercells that statistically reproduce the key pair and multisite correlation functions of a random alloy, effectively minimizing artificial periodic correlations even for small supercells of 32 atoms. The B doping was also treated within this framework by constructing SQS for ternary $\text{Si}_{1 - x - y} \text{Ge}_{x} \text{B}_{y}$ alloys. This approach enabled statistically consistent modeling of substitutional disorder of both Ge and B atoms in the Si host lattice, ensuring a realistic description of the alloy's atomic structure.

We investigated two doping regimes: (a) a low dopant concentration regime ($y_{\text{B}}=0.2$), corresponding to intentional doping levels reported in experimentally grown hexagonal-diamond samples~\cite{Fadaly2020}, and (b) a high dopant concentration regime ($y_{\text{B}}$ ranging from 1.\% to 8.2\%), consistent with experimentally achieved hyperdoping levels beyond the solubility limit~\cite{Nath2024}. In both regimes, the full range of Ge compositions was considered.

For the low concentration regime and undoped alloys (where $y_{\text{B}}=0$), we employed large supercells of $5 \times 5 \times 5$ (500 atoms) for the hexagonal phase and $4 \times 4 \times 4$ (512 atoms) for the cubic phase. A single B impurity was introduced per supercell to minimize dopant–dopant interactions between periodic images, corresponding to doping concentrations on the order of $10^{19}$ to $10^{20}$ cm$^{-3}$, approaching the experimental range of approximately $10^{18}$ cm$^{-3}$~\cite{Fadaly2020}. Structural relaxation was performed in two stages: initially, atomic positions and cell volumes were optimized for the pristine (undoped) structures; subsequently, a Si or Ge atom was substituted by B, and atomic positions were relaxed while keeping the cell volume fixed. A $2 \times 2 \times 2$ Monkhorst–Pack k-point mesh was used for both phases. Whenever feasible, substitutions at both Si and Ge sites were considered, with all relevant in-plane and out-of-plane bonding configurations taken into account. In the high dopant concentration regime, we utilized $4 \times 4 \times 4$ supercells (216 atoms) for the hexagonal phase and $3 \times 3 \times 2$ supercells for the cubic phase. Initial structures were constructed by applying Vegard’s law~\cite{DentonPRA1991} to determine lattice parameters for the SiGe alloys, followed by full relaxation of atomic positions and cell volumes under the same convergence criteria as above. A $3 \times 3 \times 3$ Monkhorst–Pack k-point grid was employed.

Given the large configurational space, to adequately sample the disorder, three independent SQS were generated for each low doping concentration, and a minimum of five for each high doping concentration (this is because larger B concentrations require more configurations). Reported results correspond to the averaged values over these configurations, with standard deviations provided to quantify the variability arising from configurational disorder.

\section{Results and discussions}
\subsection{Lattice constants} 
\begin{figure}[ht]
    \centering
    \includegraphics[width=\linewidth]{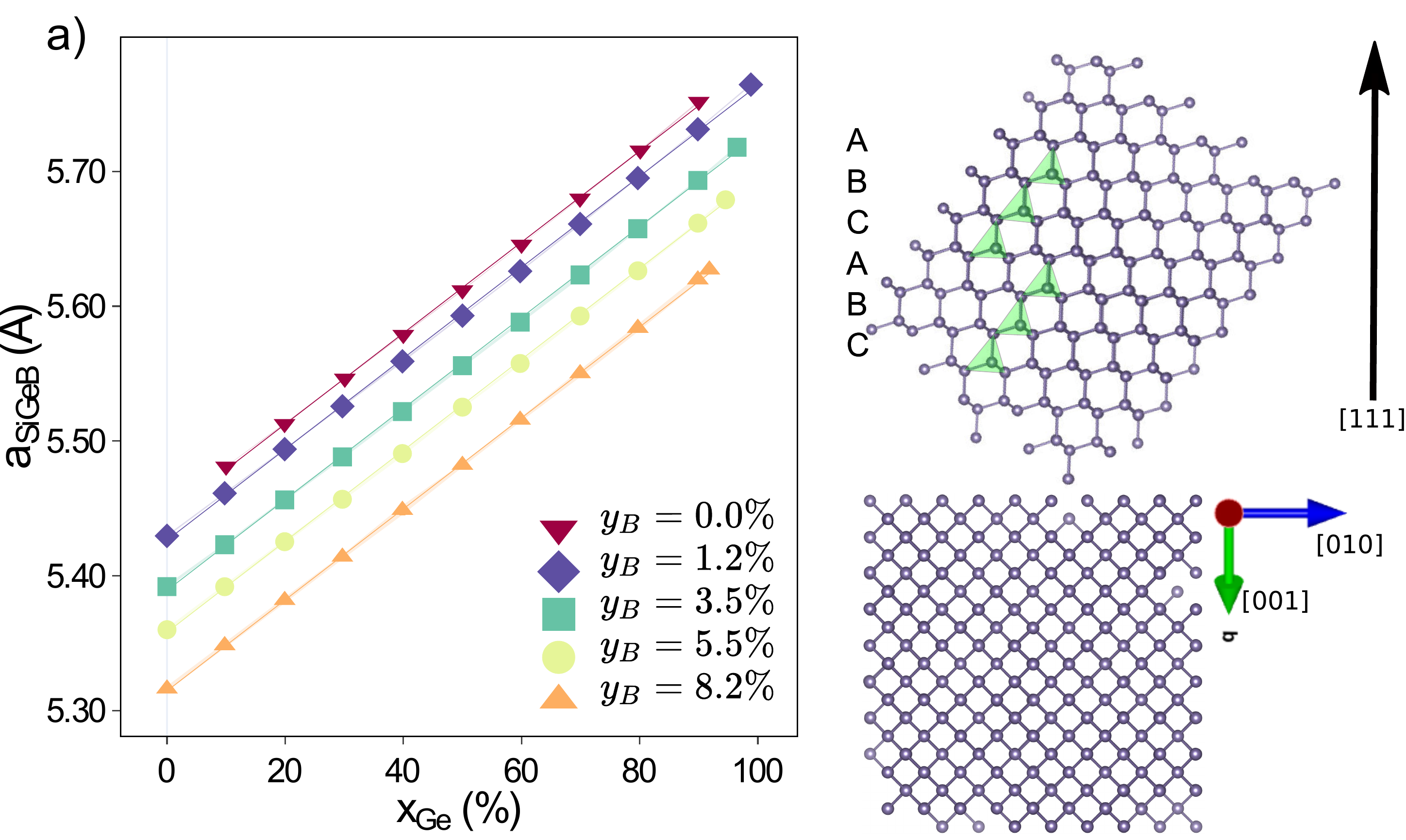}
    \caption{(a) Lattice parameter ($a_{\text{SiGeB}}$) for cub-Si$_{1-x_{\text{Ge}}-y_{\text{B}}}$Ge$_{x_{\text{Ge}}}$B$_{y_{\text{B}}}$ alloys as a function of the germanium concentration ($x_{\text{Ge}}$) for different B concentrations ($y_{\text{B}}$). Solid lines correspond to a linear fitting as in Eq.~\eqref{eq:three_vegard_law}.}
    \label{fig:figure_1}
\end{figure}

\begin{figure}[ht]
    \centering
    \includegraphics[width=\linewidth]{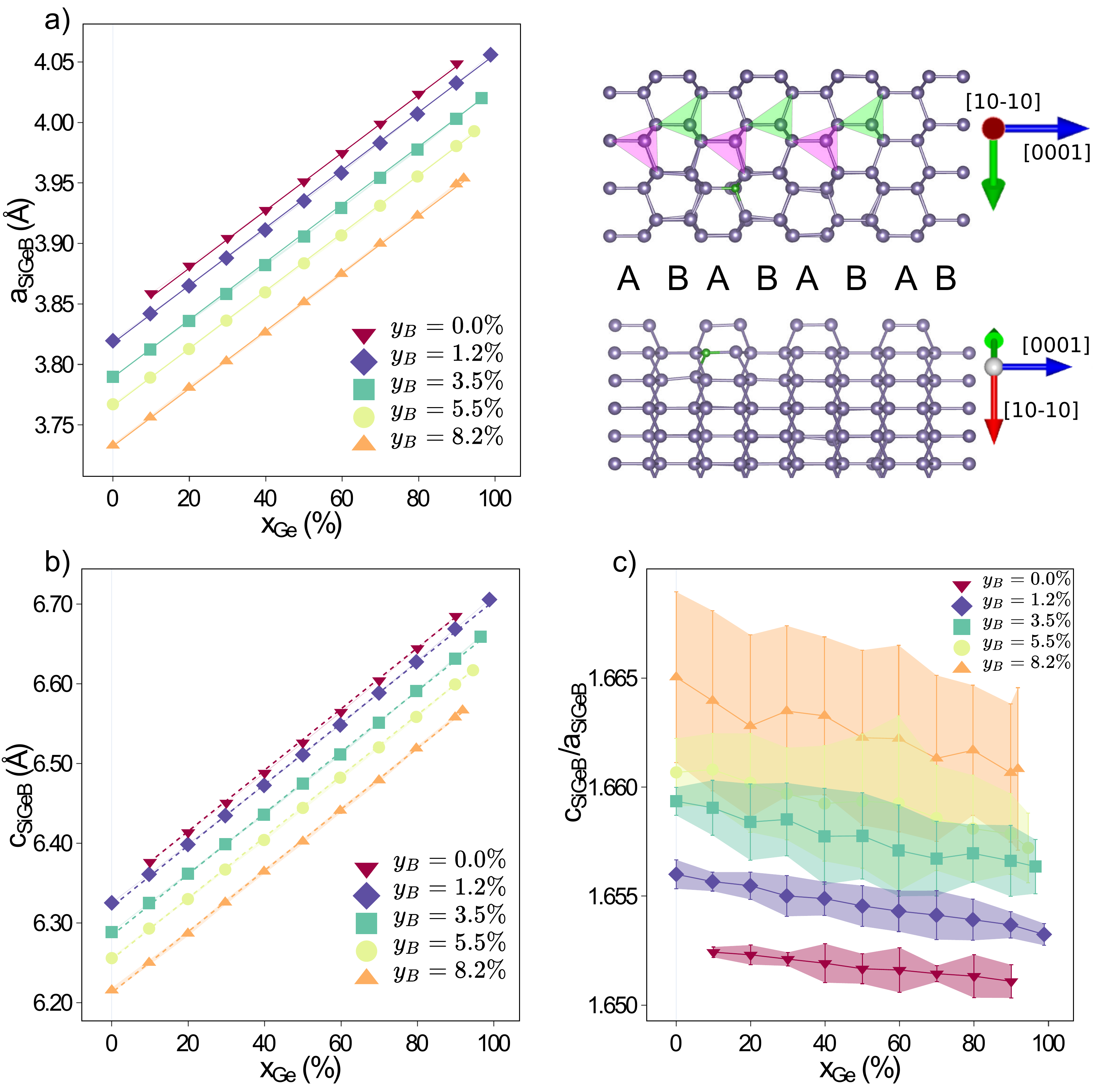}
    \caption{(a) Lattice parameter ($a_{\text{SiGeB}}$) in the in-plane of hex-Si$_{1-\text{x$_{\text{Ge}}$}-y_{\text{B}}}$Ge$_{\text{x$_{\text{Ge}}$}}$B$_{y_{\text{B}}}$ alloys as a function of the Ge concentration ($x_{\text{Ge}}$) for different B concentrations ($y_{\text{B}}$). Solid lines correspond to a linear fitting as in Eq.~\eqref{eq:three_vegard_law}. (b) Lattice parameter ($c_{\text{SiGeB}}$) in the out-of-plane of hex-Si$_{1-x_{\text{Ge}}-y_{\text{B}}}$Ge$_{x_{\text{Ge}}}$B$_{y_{\text{B}}}$ alloys as a function of the Ge concentration ($x_{\text{Ge}}$) for different B concentrations ($y_{\text{B}}$). Dashed lines correspond to a linear fitting as in Eq.~\eqref{eq:three_vegard_law}. (c) Lattice relation ($c_{\text{SiGeB}}/a_{\text{SiGeB}}$) of Si$_{1-x_{\text{Ge}}-y_{\text{B}}}$Ge$_{x_{\text{Ge}}}$B$_{y_{\text{B}}}$ alloys as a function of the Ge concentration ($x_{\text{Ge}}$) for different B concentrations ($y_{\text{B}}$). Standard deviation (see Methodology section) is included by error bars and filled area for all plots.}
    \label{fig:figure_2}
\end{figure}

Figures~\ref{fig:figure_1} and~\ref{fig:figure_2} depict the evolution of the structural parameters of SiGe alloys as functions of Ge concentration ($x_{\text{Ge}}$) and the  B concentration ($y_{\text{B}}$) for the hexagonal-diamond phase and the cubic-diamond one. In both crystal structures, a linear relationship is observed between the lattice constants of the SiGeB alloys and the concentrations of their constituent elements. This behavior, well established for binary cubic-diamond SiGe alloys, is commonly referred to as Vegard’s law~\cite{DentonPRA1991}. By fitting a ternary Vegard’s law relation (shown as continuous and dashed lines in Figures~\ref{fig:figure_2}(a) and (b), respectively), expressed as
\begin{equation}\label{eq:three_vegard_law}
    a_{\text{SiGeB}} = a_{\text{B}} y_{\text{B}} + a_{\text{Ge}}  x_{\text{Ge}} + a_{\text{Si}}  (1-x_{\text{Ge}} -y_{\text{B}} ),
\end{equation}

\noindent
we obtained the lattice parameters for the ternary alloy from DFT calculations for both phases, which are reported in the fourth and fifth columns of Table~\ref{tab:table_1}. This linear relation similarly holds for the out-of-plane lattice parameter ($c_{\text{SiGeB}}$) of the hex-SiGe alloys. The linear regression parameters show excellent agreement with the optimized GGA-PBE lattice parameters for pure hex-Si and hex-Ge (first and second columns of Table~\ref{tab:table_1}), thus confirming the validity of a three-component Vegard’s law.

In the cub-SiGeB alloys, B incorporation induces an overall lattice compression, a phenomenon recently confirmed experimentally in cub-SiGeB thin films~\cite{Nath2024}. An analogous trend is observed in the hex-SiGeB alloys, where increasing B concentration results in a contraction both within the basal plane and along the out-of-plane direction, as evidenced by the relatively small fitted lattice parameters $a_{\text{B}}$ and $c_{\text{B}}$ (fourth and fifth columns of Table~\ref{tab:table_1}). Furthermore, Figure~\ref{fig:figure_2}(c) illustrates that the $c/a$ ratio remains nearly constant across the full range of Ge concentrations (see also the third column of Table~\ref{tab:table_1}). However, increasing $y_{\text{B}}$ leads to a slight increase in the $c/a$ ratio, attributable to the intrinsically higher $c/a$ ratio of B (second row, sixth column of Table~\ref{tab:table_1}). Notably, the error bars in Figure~\ref{fig:figure_2}(c) reveal that higher B concentrations introduce more pronounced statistical deviations, reflecting structural distortions induced by B incorporation. A comparable, though less marked, effect has been reported in SiGeC alloys~\cite{Fuchs2025}, consistent with the fact that C, as a group IV element, better accommodates within the SiGe lattice than B.

\begin{figure}[ht]
    \centering
    \includegraphics[width=1.0\linewidth]{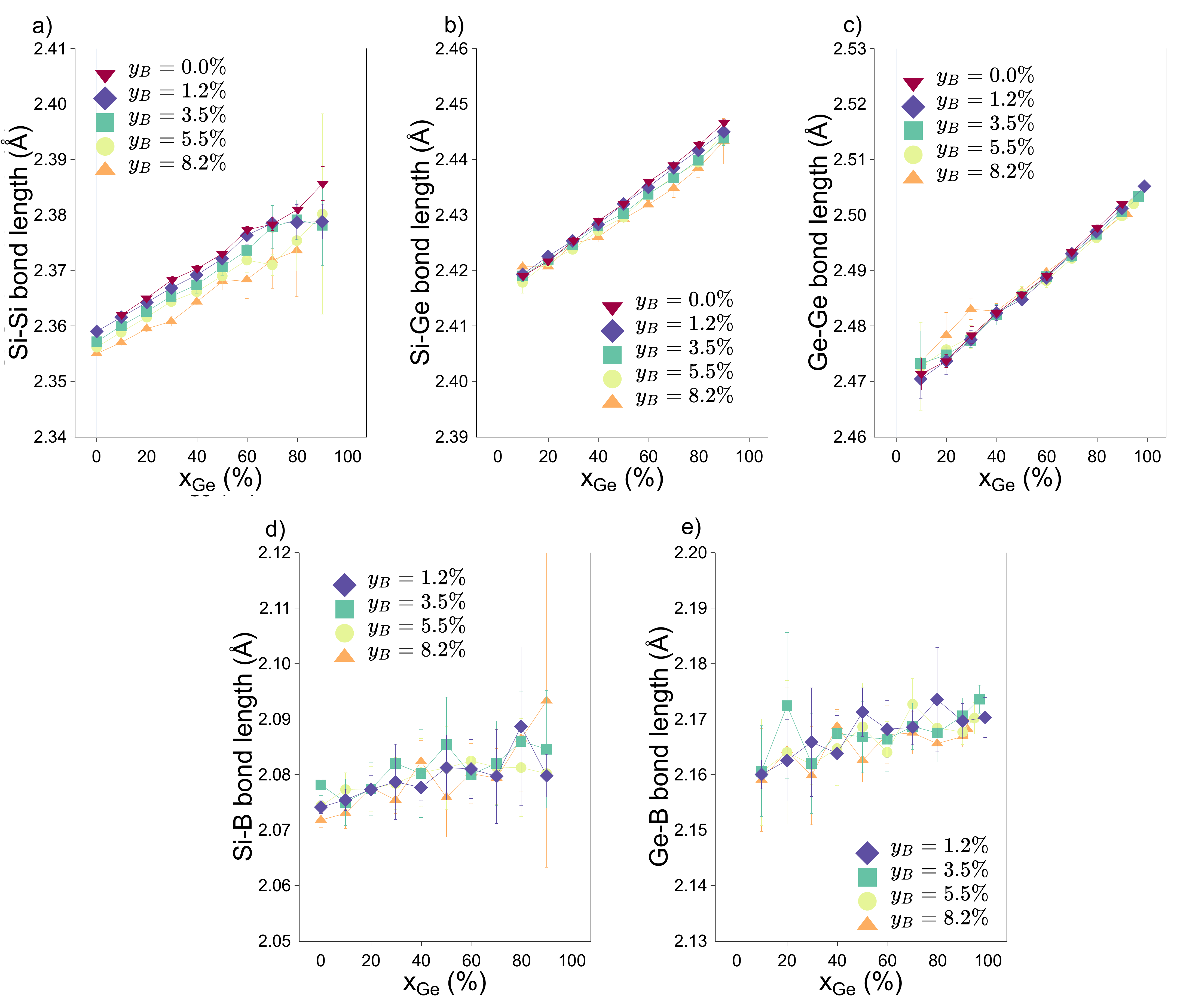}
    \caption{Bonds distances as a function of the Ge concentration between (a) Si--Si, (b) Si--Ge, (c) Ge--Ge, (d) Si--B, and (e) Ge--B for cub-Si$_{1-x_{\text{Ge}}-y_{\text{B}}}$Ge$_{x_{\text{Ge}}}$B$_{y_{\text{B}}}$. Standard deviation (see Methodology section) is included by error bars.}
    \label{fig:figure_3}
\end{figure}

\begin{figure}[ht]
    \centering
    \includegraphics[width=1.0\linewidth]{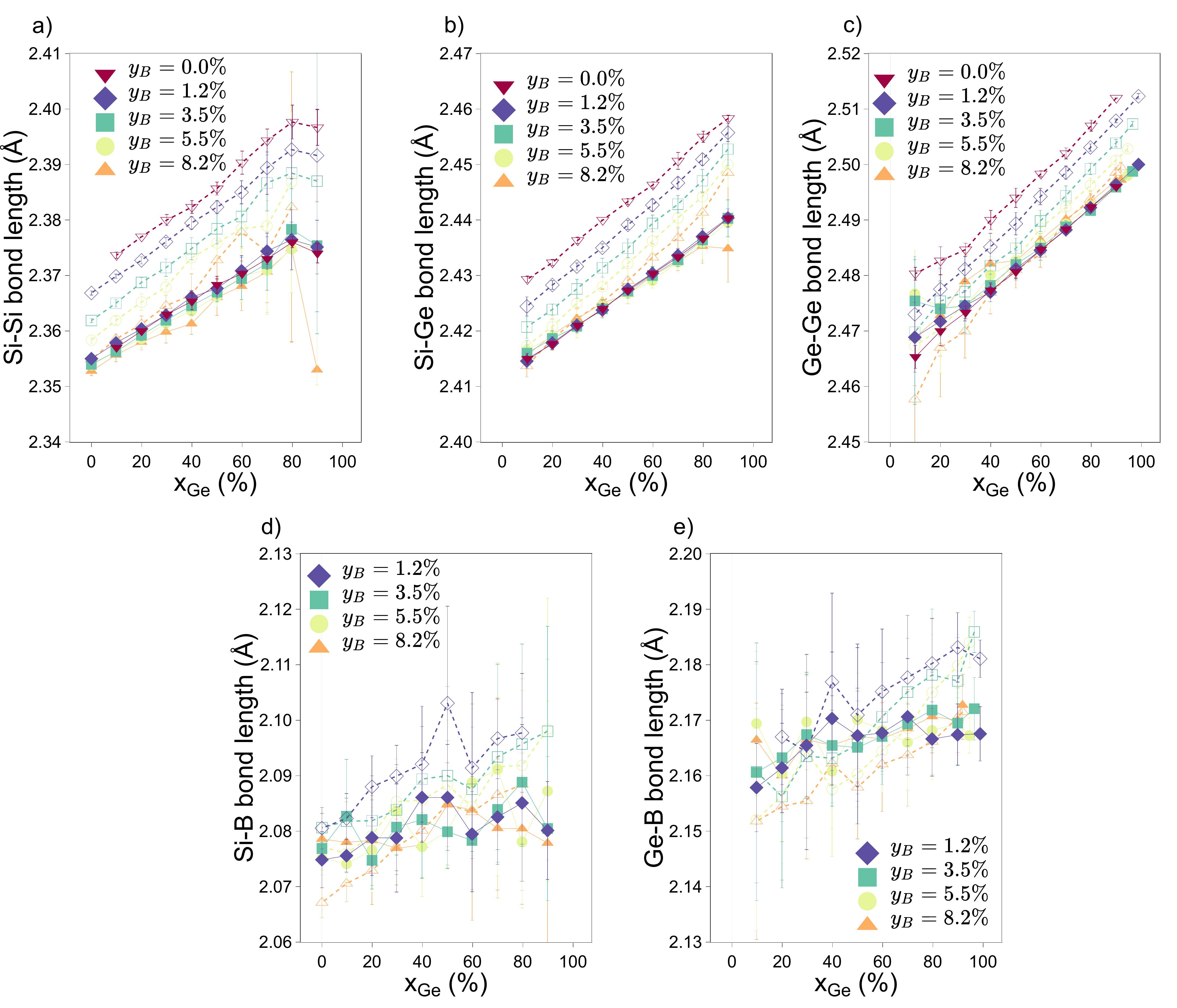}
    \caption{Bonds distances as a function of the Ge concentration between (a) Si--Si, (b) Si--Ge, (c) Ge--Ge, (d) Si--B, and (e) Ge--B in-plane (filled points and continuous lines) and out-of-plane (empty points and dashed lines) for hex-Si$_{1-x_{\text{Ge}}-y_{\text{B}}}$Ge$_{x_{\text{Ge}}}$B$_{y_{\text{B}}}$.}
    \label{fig:figure_4}
\end{figure}

A more detailed structural analysis of Si--Si, Ge--Ge, Si--Ge, Si--B, and Ge--B bonds allows us to better understand the symmetry transformations induced by doping. As shown in Figure~\ref{fig:figure_3} and~\ref{fig:figure_4}, for both phases, the bond lengths are linearly correlated with the Ge concentration. This is particularly clear for the cubic-diamond phase, where the Si--Si, Si--Ge, and Ge--Ge bond distances (see Figure~\ref{fig:figure_3} (a), (b), and (c)) linearly increase when the Ge composition increases. The effect of the B concentration, as for the lattice parameters, is to linearly reduce the bond distances. Interestingly, the local symmetry around B atoms when the B and Ge concentration varies remains of $T_{d}$ type with B forming four equal bonds with Si or Ge atoms.  

In the case of the hex-SiGeB alloys, the effect of varying the constituent composition is not isotropic as in the cubic-diamond phase because of the $C_{3v}$ lower crystal symmetry. In Figure~\ref{fig:figure_4}, dashed lines and empty points represent out-of-plane bonds, while solid lines and filled points correspond to in-plane bonds. The $C_{3v}$ bond symmetry is preserved for Si--Si, Si--Ge, and Ge--Ge at lower B concentrations, with out-of-plane bonds remaining longer than in-plane ones. However, increasing the B concentration reduces the out-of-plane bond length, making it closer to the in-plane bond length and diminishing the $C_{3v}$ symmetry. In contrast, Si-B and Ge-B bonds (Figures~\ref{fig:figure_4} (d) and (e)) do not follow a clear trend, making it complicated to identify a consistent symmetry pattern within the studied concentration range: the variations in bond lengths with $x_{\text{Ge}}$ are relatively small compared to the overall discrepancies. By comparing the length scales of Figure~\ref{fig:figure_4} (d) and (e), it is clear that Si-B bond lengths remain shorter than Ge-B bond lengths regardless of the Ge composition $x_{\text{Ge}}$, as expected due to the different covalent radius between Si and Ge. 

It is worth noting that in a few configurations for the highest B concentration, we observed B agglomeration that induced the breaking of the $C_{3v}$ symmetry, as shown in the symmetry analysis using the Ovito code~\cite{Ball2009, Maras2016} in Figure~S1 from the Supporting Information (SI). The symmetry breaks when B atoms are very close and form clusters. B-B bond distances induce the crystal symmetry change, as reported in the case of Ga-hyper doping in Ge~\cite{Steele2024}.

The influence of B concentration on the structural parameters of hex-SiGeB alloys can be summarized as follows: across the entire doping range, the lattice constants obey a ternary Vegard’s law, with the $C_{3v}$ crystal symmetry preserved at low doping concentrations. However, at elevated doping levels—specifically when $x_{\text{B}}$ exceeds approximately 5\%—a notable disruption of the $C_{3v}$ bond symmetry occurs, primarily driven by B clustering. This distortion is absent at low B concentrations (around $  y\approx 0.2\%$), where the local symmetry remains intact.

\subsection{B incorporation in the low concentration regime}
To investigate the thermodynamic behavior in the low-doping regime, we employed the Zhang-Northrup formalism, which is particularly well-suited for modeling the energetics of isolated dopants in SiGe alloy systems. Within this framework, the formation enthalpy of a substitutional B impurity in a SiGe alloy can be expressed as~\cite{Zhang1991}:
\begin{equation}\label{eq:form_energy_low}
      \Delta H_{\text{form}}  = E_{\text{SiGeB}} -  E_{\text{SiGe}} + \sum_{\text{i=Si,Ge}} n_{\text{i--j}} \hat{\mu}_{\text{i--j}} -  \mu_{\text{B}},
\end{equation}
where $E_{\text{SiGeB}}$ and $E_{\text{SiGe}}$ denote the ground-state total energies of the doped and pristine (undoped) SiGe systems, respectively. The term $\hat{\mu}_{\text{i--j}}$ represents the effective chemical potential associated with the bonds involving the substituted atom, obtained by summing over the contributions from all relevant bond types (Si–Si, Ge–Ge, and Si–Ge). Within the quasi-chemical model approximation, the total internal energy of the SiGe alloy is only related to the bond energies between adjacent atoms. The quantity $\hat{\mu}_{\text{i--j}}$ can be hence derived by decomposing the DFT total energy of the undoped SiGe alloy, $E_{\text{SiGe}}$, into the corresponding bond energies, reflecting the local bonding environment of the substituted site:
\begin{equation}\label{eq:fitted_energy}
    E_{\text{SiGe}} (n_{\text{Si--Si}}, n_{\text{Ge--Ge}}, n_{\text{Si--Ge}})  = n_{\text{Si--Si}} \hat{\mu}^{\text{fit}}_{\text{Si--Si}} +n_{\text{Si--Ge}} \\
    \hat{\mu}^{\text{fit}}_{\text{Si--Ge}} + n_{\text{Ge--Ge}} \hat{\mu}^{\text{fit}}_{\text{Ge--Ge}}
\end{equation} 
\noindent
where $n_{\text{i--j}}$ is the number of bonds between species $\text{i}$ and $\text{j}$ and  $\hat{\mu}^{\text{fit}}_{\text{i--j}}$ the energy of the bond between the species $i$ and $j$. By fitting the DFT total energies as a function of the number of bonds for the undoped system, we obtained that $\hat{\mu}^{\text{fit}}_{\text{Si--Si}} \approx \mu_{\text{Si}} / 2 $ and  $\hat{\mu}^{\text{fit}}_{\text{Ge--Ge}} \approx \mu_{\text{Ge}} / 2 $, where $\mu_{\text{i}}$ is the ground-state total energy per atom of the bulk, with an error of around 10~meV, and when bonds are counted with multiplicity 1. On the other hand, $\hat{\mu}^{\text{fit}}_{\text{SiGe}}$ lies between $\hat{\mu}^{\text{fit}}_{\text{Si--Si}}$ and $\hat{\mu}^{\text{fit}}_{\text{Ge--Ge}}$ (Table~S1 from the SI). 

The fourth term of Eq.~\ref{eq:form_energy_low} is the B chemical potential, $\mu_{\text{B}}$, taken as the ground-state total energy per atom of the most stable allotrope of B under normal conditions, the $\beta$-rhombohedral structure (R$\bar{3}$m space group)~\cite{Hayami2024, Hoard1970}, which approximates the energy of an isolated impurity (Table~S1 from the SI). It is worth noting that the determination of the B chemical potential is quite arbitrary because it strongly depends on the synthesis conditions. However, as our focus is on comparing the formation enthalpy difference of B between the hexagonal and cubic phases, and given that \(\mu_{\text{B}}\) is assumed to be the same in both, this choice does not impact the physical conclusions.

\begin{figure}[ht]
    \centering
    \includegraphics[width=0.6\linewidth]{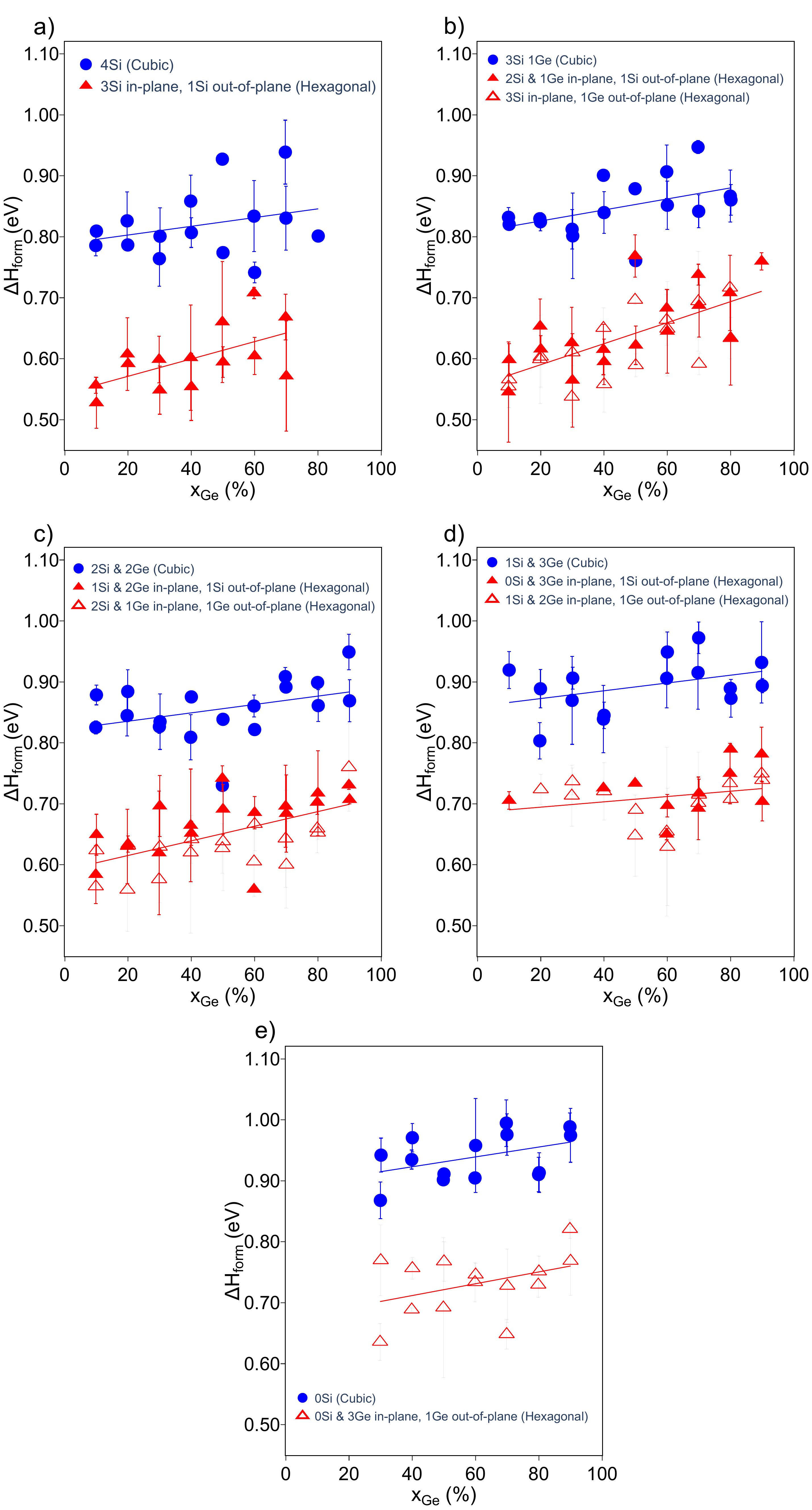}
    \caption{Formation energy, as expressed in Eq.~\eqref{eq:form_energy_low}, as a function of the Ge concentration for different numbers of Si, Ge bonds, and a concentration with $y_{\mathrm{B}} = 0.2\%$. (a) For 4 Si-B bonds. (b) For 3 Si-B and 1 Ge-B. (c) For 2 Si-B and 2 Ge-B. (d) For 1 Si-B and 3 Ge-B. (e) For 4 Ge-B bonds. The hexagonal phase (red triangles) and cubic phase (blue circles) are included. For the hexagonal phase (red triangles), empty points correspond to a Ge line in the out-of-plane bonds, and filled points correspond to a Si atom in the out-of-plane bond. Standard deviation (see Methodology section) is included in the error bars.}
    \label{fig:figure_5}
\end{figure}

Figure~\ref{fig:figure_5} shows the formation energy of a B atom as a function of Ge concentration in both the hexagonal and cubic phases, considering all possible bond configurations. The panels are organized according to the number of Si–B bonds. The cases where either a Si or Ge atom is substituted are shown without distinction, as they yield comparable formation energies. Empty red markers indicate configurations where the out-of-plane neighbor is a Ge atom, while filled markers correspond to an out-of-plane Si atom. In all bond configurations, the formation energy of B is consistently lower in the hex-SiGe phase compared to the cubic one, with differences ranging from 10~meV to 200~meV. This trend is consistent with previous results reported for B in pure Si and Ge~\cite{Amato2019, Amato2020, Tunica2024}. Furthermore, a comparison across panels reveals that configurations involving a greater number of Ge atoms are energetically slightly less favorable, with an increase in formation energy of approximately 10~meV, regardless of whether the Ge atom is located in-plane or out-of-plane.

Importantly, in the low-concentration regime, B atoms exhibit a clear energetic preference for incorporation into the hex-SiGe phase. This suggests that $p$-type doping of the hexagonal phase should be thermodynamically favored. Since successful doping of the cubic SiGe phase has already been demonstrated experimentally~\cite{Nath2024}, this result offers strong promise that controlled hole doping of hexagonal SiGe alloys could also be realized in practice.

\subsection{B incorporation in the high concentration regime}
In the high-concentration regime, we adopt a mixing enthalpy approach, which is more suitable than the Zhang–Northrup formalism for this context, as the system can be effectively described as a solid solution due to the substantial number of B substitutions~\cite{Bustarret2006}. The evaluation of the mixing enthalpy enables the assessment of the thermodynamic stability of the alloy relative to its pure elemental constituents~\cite{Borlido2021,Fuchs2025}. It is computed by subtracting the total energy of the ideal solid solution (where the interactions between atoms Si, Ge, and B are identical, and the enthalpy variation due to the mixing is zero):
\begin{equation}\label{eq:ideal_enthalpy}
    E_{\text{SiGeB}}^{\text{ideal}}  =    N_{\text{Si}}  \mu_{\text{Si}} + N_{\text{Ge}}  \mu_{\text{Ge}} + N_{\text{B}}  \mu_{\text{B}}  
\end{equation}  
\noindent
from the DFT total energy of the real alloy (where the energy of the elastic strain fields due to the
mismatch in atomic sizes is taken into account), $E_{\text{SiGeB}}$, normalized per atom, as in the following equation: 
\begin{equation}\label{eq:mixing_enthalpy}
    \Delta H_{\text{mix}}  =   \frac{1}{N}  \left[ E_{\text{SiGeB}}  - E_{\text{SiGeB}}^{\text{ideal}} \right]  
\end{equation} 
\noindent
where $N$ = $N_{\text{Si}}$ + $N_{\text{Ge}}$ + $N_\text{B}$ is the total number of atoms in the supercell and $\mu_i$ is the chemical potential of each atomic species (Si, Ge, and B) as defined in the previous section.

\begin{figure}[ht]
    \centering
    \includegraphics[width=0.8\linewidth]{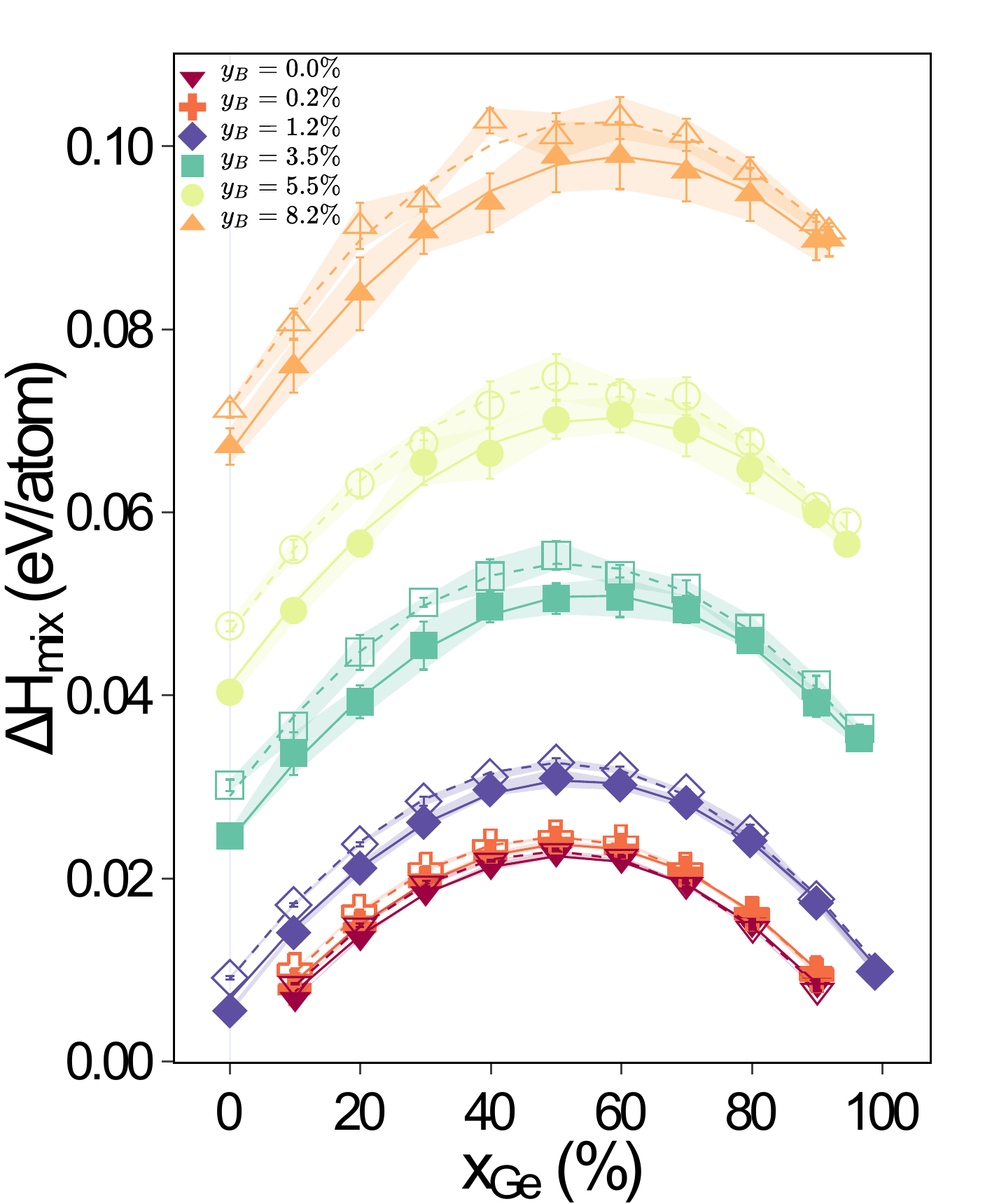}
    \caption{Mixing enthalpy as calculated in Eq.~\eqref{eq:mixing_enthalpy} as a function of the Ge concentration for different B concentrations. The colored area corresponds to the standard deviation. Lines correspond to the fitting (see Table~\ref{tab:h_parameters}) of Eq.~\eqref{eq:equation_assymetric_non_linear} for hex- (solid line) and cub-$Si_{1-x_{\text{Ge}}-y_{\text{B}}}Ge_{x_{\text{Ge}}}B_{y_{\text{B}}}$ (dashed line).}
    \label{fig:figure_6}
\end{figure}

Figure~\ref{fig:figure_6} presents the average mixing enthalpy, \( \Delta H_{\text{mix}} \), of $\text{Si}_{1 - x - y} \text{Ge}_{x} \text{B}_{y}$ as a function of $x_{\text{Ge}}$, for various B concentrations $y_{\text{B}}$. Both hexagonal-diamond (solid lines) and cubic-diamond (dashed lines) alloy structures are considered. For \( y_{\text{B}}= 0 \), the results agree with previous theoretical studies on hex-SiGe alloys~\cite{Borlido2021}, confirming the validity of our approach. In all cases, the mixing enthalpies are positive, with the lowest values observed for \( y_{\text{B}} = 0 \). The value of \( \Delta H_{\text{mix}} \) increases with \( y_{\text{B}} \) for any given \( x_{\text{Ge}} \), indicating that B incorporation is slightly energetically unfavorable. However, the enthalpy increase is relatively small and can likely be overcome at room temperature. The increase in mixing enthalpy relative to the undoped case, \( \Delta H_{\text{mix}}(y) - \Delta H_{\text{mix}}(0) \), becomes larger at higher B concentrations, suggesting that the energy cost of B incorporation becomes more significant with increasing \( y_{\text{B}}\). This behavior deviates notably from linearity, pointing to complex interactions at elevated doping levels.

For a fixed $y_{\text{B}}$, the enthalpy mixing energy reaches its maximum at approximately $x\approx$50~\%, also for the undoped material ($y_{\text{B}} = 0$). This behavior is driven by the energy cost due to the creation of the Si--Ge bonds, which require more energy than Si-Si and Ge--Ge bonds. At low B concentrations, the mixing enthalpy profiles of the two phases are nearly identical. However, with increasing B content, the hex-SiGeB phase becomes progressively more favorable. This is a noteworthy result, as it suggests that B hyperdoping of the hex-SiGe phase may, in principle, be thermodynamically more favorable than hyperdoping the cub-phase.

The difference in enthalpy, $\Delta H_{\text{mix}}$, can be approximately described using a regular solution model generalized for a ternary alloy system~\cite{Windl2022}:

\begin{equation}\label{eq:equation_assymetric_non_linear}
\begin{split}
\Delta H_{\text{mix}} (x_{\text{Ge}}, y_{\text{B}}) \approx &~ 
\left[ H_0^{\text{SiGe}} (1 + H_1^{\text{SiGe}}(1 - x_{\text{Ge}}- y_{\text{B}})) \right] x_{\text{Ge}}(1 -  x_{\text{Ge}} - y_{\text{B}}) \\ 
& + \left[ H_0^{\text{GeB}} (1 + H_1^{\text{GeB}} y_{\text{B}}) \right] x_{\text{Ge}} y_{\text{B}}  \\ 
& + \left[ H_0^{\text{SiB}} (1 + H_1^{\text{SiB}}(1 - x_{\text{Ge}} - y_{\text{B}})) \right] y_{\text{B}}(1 - x_{\text{Ge}} - y_{\text{B}})  \\ 
& + \left[ H_0^{\text{BB}} (1 + H_1^{\text{BB}} y_{\text{B}}) \right] y_{\text{B}}^2, 
\end{split}
\end{equation}

\noindent
where $H^{\text{ij}}_{0}$ and $H^{\text{ij}}_{1}$ are the pairwise interaction parameters between components $i$ and $j$, capturing symmetric and asymmetric mixing contributions, respectively. The B--B interaction term is included because our energy reference is a B allotrope different from the hexagonal-diamond phase. 

The pairwise interaction parameters are obtained by fitting the $\Delta H_{\text{mix}}$ calculated with DFT as a function of the $x_{\text{Ge}}$ and $y_{\text{B}}$ with Eq.~\eqref{eq:equation_assymetric_non_linear} (see Figure~\ref{fig:figure_6}). The resulting values are reported in Table~\ref{tab:h_parameters}, and they support the previous thermodynamic stability arguments. The $H_0^{\text{SiGe}}$ values are small and similar for both crystal structures, indicating that the Si–Ge bonding character is largely unaffected by the underlying lattice symmetry. This suggests that Si–Ge alloys are highly miscible in both cubic and hexagonal phases. In contrast, $H_0^{\text{GeB}}$ and $H_0^{\text{SiB}}$ are significantly larger (on the order of hundreds of meV), reflecting a much stronger enthalpy cost for mixing B with either Ge or Si. These parameters are lower in the hexagonal phase, implying that the hexagonal lattice may better accommodate B atoms. Among the two, B--Ge interaction is stronger than B–Si in both crystal structures, particularly in the cubic phase. Finally, $H_0^{\text{BB}}$ is very large (4–6 eV), indicating strong repulsion or instability of B–B pairs in both structures. Interestingly, B–B interaction parameters are lower in the cubic phase, meaning that B–B clustering is even more energetically unfavorable in the hexagonal structure. The large and positive value of $H_{1}^{\text{GeB}}$ calculated for the cubic phase may further complicate the mixing behavior, potentially leading to bowing or phase instability that could be somewhat mitigated in the hexagonal phase. Overall, B incorporation is highly penalized energetically—especially in the cubic phase and particularly for B–B clustering. While the hexagonal phase may slightly improve B solubility, the overall solubility remains limited by the high enthalpy cost.

It is important to note that Figure~\ref{fig:figure_6} alone does not allow us to distinguish between the contributions to $\Delta H_{\text{mix}}$ from B incorporation and those arising from Si--Ge bonding. Indeed, since the approach in Eq.~\ref{eq:mixing_enthalpy} is based on an average over all atoms in the system, it is insightful to shift our attention to the bond energies within the alloys involving B atoms. This perspective is particularly useful for investigating solubility and stability in ternary systems, especially when compared to the corresponding binary SiGe alloys. We obtained a rough estimate of the pairwise interaction parameters, $H_{0}^{\text{ij}}$ and $H_{1}^{\text{ij}}$, by fitting the dependence of $\Delta H_{\text{mix}}$ on $x_{\text{Ge}}$ and $y_{\text{B}}$ in Eq.~\eqref{eq:equation_assymetric_non_linear}. However, such a fit involves many parameters, making it complex and not highly accurate. 

To further confirm our previous results, we can try to isolate the variation of enthalpy of introducing B into the SiGe alloys, $\Delta H_{\text{mix}}^B$ by using the exact SiGe bond energy value, $\hat \mu_{\text{SiGe}}^{\text{fit}}$ taken from Eq.~\eqref{eq:fitted_energy} and rewriting an expression for $E_{\text{SiGeB}}^{\text{ideal}}$. Let’s start by considering Eq.~\eqref{eq:ideal_enthalpy}, which expresses the total energy of the ideal solid solution. It can be rewritten as a function of the number of bonds as
\begin{equation}\label{eq:ideal_enthalpy_bonds}
\begin{split}
    E_{\text{SiGeB}}^{\text{ideal}}  =~ &    n_{\text{Si--Si}} \, \hat{\mu}^{\text{avg}}_{\text{Si--Si}} 
- n_{\text{Ge--Ge}} \, \hat{\mu}^{\text{avg}}_{\text{Ge--Ge}}  - n_{\text{Si--Ge}} \, \hat{\mu}^{\text{avg}}_{\text{Si--Ge}} \\
& - n_{\text{Si--B}} \, \hat{\mu}^{\text{avg}}_{\text{Si--B}} 
- n_{\text{Ge--B}} \, \hat{\mu}^{\text{avg}}_{\text{Ge--B}} 
- n_{\text{B--B}} \, \hat{\mu}^{\text{avg}}_{\text{B--B}}
\end{split}
\end{equation}  
where \( \hat{\mu}^{\text{avg}}_{\text{i--j}} \) is the bond energy between species $i$ and $j$ calculated as the average of the chemical potentials of each species, $\mu_{\text{i}}$, i.e. \( \hat{\mu}^{\text{avg}}_{\text{i--j}} = (\mu_{\text{i}} + \mu_{\text{j}})/4 \). Then, it is straightforward to see that Eq.~\eqref{eq:ideal_enthalpy} and Eq.~\eqref{eq:ideal_enthalpy_bonds} are equivalent. 

\begin{figure}[ht]
    \centering
    \includegraphics[width=0.9\linewidth]{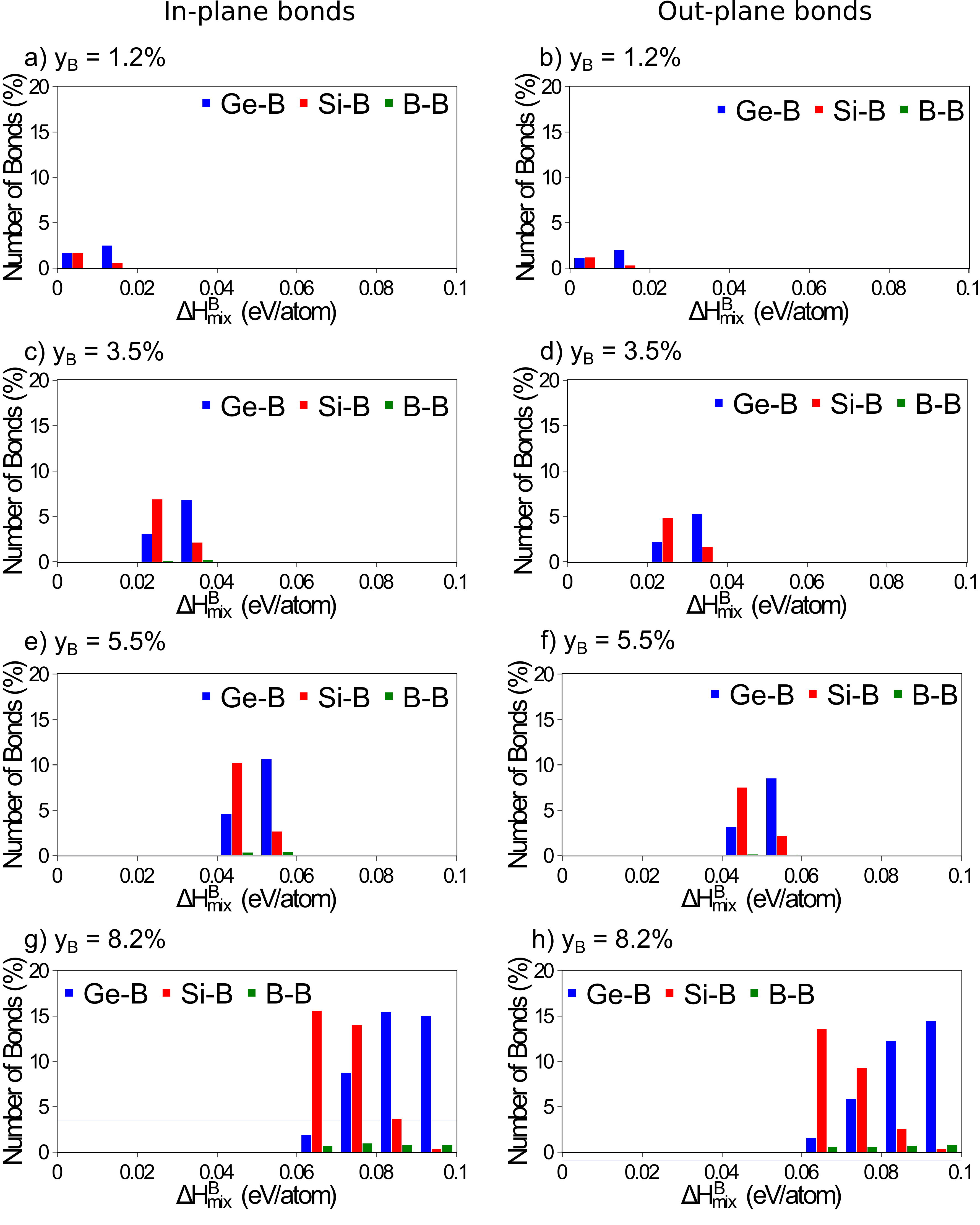}
    \caption{Histogram of the number of Si-B, Ge-B, and B-B bonds divided by each bin's total number of bonds as a function of the excess energy as in Eq.~\eqref{eq:excess_energy_B}. Each row corresponds to a different B concentration. The column on the left corresponds to the in-plane bond and the one on the right to the out-of-plane bond.}
    \label{fig:figure_7}
\end{figure}

In Eq.~\eqref{eq:fitted_energy} , the first three terms can be replaced by $E_{\text{SiGe}}$ of Eq.~\eqref{eq:fitted_energy}, to take into account the exact Si--Ge bond energy, $\hat \mu_{\text{SiGe}}^{\text{fit}}$ instead of using $\hat{\mu}^{\text{avg}}_{\text{Si--Ge}}$:
\begin{equation}\label{eq:ideal_enthalpy_B}
\begin{split}
E_{\text{SiGeB}}^{\text{ideal B}} = & E_{\text{SiGe}} + n_{\text{Si--B}} \, \hat{\mu}^{\text{avg}}_{\text{Si--B}} 
+ n_{\text{Ge--B}} \, \hat{\mu}^{\text{avg}}_{\text{Ge--B}} 
+ n_{\text{B--B}} \, \hat{\mu}^{\text{avg}}_{\text{B--B}} \\
= & \, n_{\text{Si--Si}} \, \hat{\mu}^{\text{fit}}_{\text{Si--Si}} 
+ n_{\text{Ge--Ge}} \, \hat{\mu}^{\text{fit}}_{\text{Ge--Ge}} 
+ n_{\text{Si--Ge}} \, \hat{\mu}^{\text{fit}}_{\text{Si--Ge}} \\
&+ n_{\text{Si--B}} \, \hat{\mu}^{\text{avg}}_{\text{Si--B}} 
+ n_{\text{Ge--B}} \, \hat{\mu}^{\text{avg}}_{\text{Ge--B}} 
+ n_{\text{B--B}} \, \hat{\mu}^{\text{avg}}_{\text{B--B}}.
\end{split}
\end{equation}
\noindent Notice that as discussed for Eq.~\eqref{eq:form_energy_low}, $\hat{\mu}^{\text{fit}}_{\text{Si--Si}} \approx \hat{\mu}^{\text{avg}}_{\text{Si--Si}}$ and $\hat{\mu}^{\text{fit}}_{\text{Ge--Ge}} \approx \hat{\mu}^{\text{avg}}_{\text{Ge--Ge}}$. Thus, the only difference between Eq.~\eqref{eq:ideal_enthalpy} and Eq.~\eqref{eq:ideal_enthalpy_B} lies in the energy of the Si--Ge bond. By making the difference between Eq.~\eqref{eq:ideal_enthalpy_B} and \eqref{eq:ideal_enthalpy_bonds}, we obtain
\begin{equation}\label{difference_ideal}
E_{\text{SiGeB}}^{\text{ideal B}} - E_{\text{SiGeB}}^{\text{ideal}} \approx  n_{\text{Si--Ge}} \left(\, \hat{\mu}^{\text{fit}}_{\text{Si--Ge}} - \, \hat{\mu}^{\text{avg}}_{\text{Si--Ge}} \right).
\end{equation}
\noindent
With this clarification, we can define the enthalpy of mixing due to the B incorporation into the alloy, ($\Delta H^{\text{B}}_{\text{mix}}$), in a similar way to Eq.~\eqref{eq:mixing_enthalpy}:
\begin{equation}\label{eq:excess_energy_B} 
\begin{split}
\Delta H^{\text{B}}_{\text{mix}} = & \frac{1}{N} \Big[ E_{\text{SiGeB}} -  E_{\text{SiGeB}}^{\text{ideal B}}\Big] \\
= & \frac{1}{N} \Big[ \, E_{\text{SiGeB}}
- n_{\text{Si--Si}} \, \hat{\mu}^{\text{fit}}_{\text{Si--Si}} 
- n_{\text{Ge--Ge}} \, \hat{\mu}^{\text{fit}}_{\text{Ge--Ge}} - n_{\text{Si--Ge}} \, \hat{\mu}^{\text{fit}}_{\text{Si--Ge}} \\
& 
- n_{\text{Si--B}} \, \hat{\mu}^{\text{avg}}_{\text{Si--B}} 
- n_{\text{Ge--B}} \, \hat{\mu}^{\text{avg}}_{\text{Ge--B}} 
- n_{\text{B--B}} \, \hat{\mu}^{\text{avg}}_{\text{B--B}} \Big].
\end{split}
\end{equation}
\noindent
In Figure~S2 of the SI, we report its dependence on \( x_{\text{Ge}} \) and \( y_{\text{B}} \). This quantity contains only the difference in the formation of B--B, Si--B, and Ge--B bonds with respect to the pure components, as it is zero when \( y_B= 0 \). The trends observed in Figure~\ref{fig:figure_6} persist, with enthalpy differences between various B concentration values remaining of the same order of magnitude. However, unlike the parabolic dependence of \( \Delta H_{\text{mix}} \) on \( x\), which shows a maximum at 50\%, \( \Delta H_{\text{mix}}^{\text{B}} \) follows a nearly linear trend with respect to the B concentration. Moreover, it slightly increases at high Ge concentrations, likely due to the formation of additional Ge--B bonds, which are energetically more costly than Si--B bonds. This interpretation is further supported by the bond statistics shown in Figure~\ref{fig:figure_7}, where we plot histograms of the number of bonds as a function of the \( \Delta H_{\text{mix}}^{\text{B}} \) (Eq.~\eqref{eq:excess_energy_B}) for different B concentrations. From this figure, it is evident that increasing the number of Si--B and Ge--B bonds (i.e., increasing the B composition \( y_{\text{B}} \)) leads to higher values of \( \Delta H_{\text{mix}}^{\text{B}} \). Furthermore, for all B concentrations, configurations with a larger number of Ge--B bonds exhibit higher  \(\Delta H_{\text{mix}}^{\text{B}} \) compared to those with more Si--B bonds, regardless of whether the bond is in-plane (left column) or out-of-plane (right column). This analysis confirms that the total mixing enthalpy of SiGeB alloys exhibits a parabolic dependence on \( x_{\text{Ge}} \) while varying approximately linearly with \( y_{\text{B}} \).

\section{Conclusions}
In conclusion, this work constitutes the first comprehensive theoretical study of B stability in hex-SiGe alloys. Using density functional theory simulations combined with special quasirandom structures, we systematically investigated the structural and thermodynamic stability of hexagonal-diamond B-doped SiGe alloys over the full range of Ge concentrations, considering both low and high levels of B doping and comparing them with the case of cub-SiGeB alloys. An analysis of the structural properties shows that hex-$\text{Si}_{1 - x - y} \text{Ge}_{x} \text{B}_{y}$ alloys approximately follow a ternary Vegard’s law. However, at high B concentrations, local symmetry breaking can occur, particularly in the presence of B clustering. Total energy calculations demonstrate that B incorporation is thermodynamically favored in the hex-phase relative to the cubic phase for all Ge concentrations and doping regimes. This energetic preference is relevant in the context of potential superconducting behavior. Indeed, superconductivity was theoretically predicted and experimentally demonstrated in cub-SiGe under ultra-high B doping~\cite{Nath2024,Blase2009}. Ultra-doping is instrumental for superconductivity to achieve a large enough electron-phonon coupling, increasing the density of states thanks to the larger carrier density, and promoting phonon softening through the doping-induced strain. In addition, past works have shown that superconductivity is often found in both the cubic phase and hexagonal polytypes of boron ultra-doped Si and SiC~\cite{KrienerPRB2008,Mignot1986}. The results of this work, providing insights into the role of B in stabilizing the hexagonal phase, show that ultra-doping a hexagonal SiGe phase should be experimentally achievable, even more easily than in the cubic phase, where superconductivity was already demonstrated. Ultra-doped SiGe appears thus as an ideal system, where a tuneable direct band gap could be implemented in superconducting devices in an emerging group IV, CMOS-compatible, quantum technology platform.\\

\section*{Acknowledgments}
M.A. and M.T. acknowledge the ANR AMPHORE project (ANR-21-CE09-0007) and the ANR TULIP (ANR-24-CE09-5076). F.C. acknowledges the ANR project SUNISIDEUP (ANR-19-CE47-0010-03) and the ANR project TOBITS (ANR-22-QUA2-0002-02). Part of the high-performance computing resources for this project were granted by the Institut du développement et des ressources en informatique scientifique (IDRIS) under the allocations AD010914974 and AD010915077 via GENCI (Grand Equipement National de Calcul Intensif).

\clearpage

\bibliographystyle{unsrt}  
\bibliography{manuscript}

\pagebreak
\newpage

\pagebreak
\newpage

\pagebreak
\newpage

\begin{table}[ht]
    \caption{The first three columns correspond to values directly obtained from DFT calculations for the bulk materials (experimental values are given in parentheses). The last three columns show the results of the fit based on the DFT data through Eq.~\eqref{eq:three_vegard_law}.}
    \vspace{5mm}
    \label{tab:table_1}
    \centering
    \begin{tabular}{c|c|c|c|c|c|c}
        Material  & $a_{\text{GGA}}$ (Å) & $c_{\text{GGA}}$ (Å) & $(c/a)_{\text{GGA}}$ & $a_{\text{fit}}$ (Å) & $c_{\text{fit}}$ (Å) & $(a/c)_{\text{fit}}$ \\ \hline
        Hex-Si  & 3.84 (3.82--3.84$^a$)  & 6.34 (6.26--6.34$^a$) & 1.65   & 3.83  & 6.34   & 1.66   \\
        Hex-Ge  & 4.07 (3.98$^{b}$) & 6.73 (6.57$^{b}$) & 1.65   & 4.07   & 6.72   & 1.65    \\
        Hex-B   & -  & - & - & 2.62& 4.80& 1.83\\
        Cub-Si  & 5.45$^c$ (5.43$^d$)  & - & - & 5.44& - & -      \\
        Cub-Ge  & 5.79$^c$ (5.65$^e$) & - & - & 5.78& - & -      \\
        Cub-B   & - & - & - & 3.86& - & -      \\ \hline
        \multicolumn{7}{l}{\footnotesize $^a$ From Ref.~\cite{Keller2023}, \cite{Hauge2015}, \cite{Zhang1999}, \cite{Ahn2021}, \cite{Besson1987}.} \\
        \multicolumn{7}{l}{ \footnotesize $^b$ From Ref.~\cite{Fadaly2020}.} \\
        \multicolumn{7}{l}{ \footnotesize $^c$ From Ref.~\cite{Nath2024}.} \\
         \multicolumn{7}{l}{ \footnotesize $^d$ From Ref.~\cite{Isherwood1966}, \cite{Godwod1974}.}\\
         \multicolumn{7}{l}{ \footnotesize $^e$ From Ref.~\cite{Hu2003}.} \\
   \end{tabular}
\end{table}

\pagebreak
\newpage

\pagebreak
\newpage

\pagebreak
\newpage

\pagebreak
\newpage

\pagebreak
\newpage

\begin{table}[ht]
\centering
\caption{Comparison of fitted $H$ coefficients of Eq.~\eqref{eq:equation_assymetric_non_linear} for hex- and cub-$Si_{1-x_{\text{Ge}}-y_{\text{B}}}Ge_{x_{\text{Ge}}}B_{y_{\text{B}}}$ alloys.}

\vspace{10mm}
\renewcommand{\arraystretch}{1.2}
\begin{tabular}{c|c|c}
\hline
\textbf{Coefficient} & \textbf{Hexagonal} & \textbf{Cubic} \\
\hline
$H_0^{\text{SiGe}}$ & 95.66 meV   & 92.63 meV   \\
$H_1^{\text{SiGe}}$ & -0.11       & -0.02       \\
$H_0^{\text{GeB}}$  & 843.93 meV  & 884.86 meV  \\
$H_1^{\text{GeB}}$  & -0.39       & 1.49        \\
$H_0^{\text{SiB}}$  & 536.75 meV  & 808.79 meV  \\
$H_1^{\text{SiB}}$  & -0.03       & -0.09       \\
$H_0^{\text{BB}}$   & 6083.28 meV & 3986.30 meV \\
$H_1^{\text{BB}}$   & -4.09       & -5.01       \\
\hline
\end{tabular}
\label{tab:h_parameters}
\end{table}

\pagebreak
\newpage

\end{document}